# Quantum Circuits for Dynamic Runtime Assertions in Quantum Computation

Huiyang Zhou and Gregory T. Byrd
Dept. of Electrical and Computer Engineering, North Carolina State University

**Abstract**—In this paper, we propose quantum circuits for runtime assertions, which can be used for both software debugging and error detection. Runtime assertion is challenging in quantum computing for two key reasons. First, a quantum bit (qubit) cannot be copied, which is known as the non-cloning theorem. Second, when a qubit is measured, its superposition state collapses into a classical state, losing the inherent parallel information. In this paper, we overcome these challenges with runtime computation through ancilla qubits, which are used to indirectly collect the information of the qubits of interest. We design quantum circuits to assert classical states, entanglement, and superposition states.

## 1 INTRODUCTION

Quantum computing features unique advantages over classical computing and recent advances in quantum computer hardware raise high hopes to realize the remarkable potential of quantum computing. However, developing quantum programs remains difficult, and debugging them is also highly challenging. The prior work by Huang et al. [2] shows that many bugs in quantum programs can be detected using assertions. Assertions, especially dynamic ones, during quantum program execution are challenging for two key reasons. The first is the non-cloning theorem, which means that it is impossible to copy a quantum bit (qubit) with an arbitrary state. The second is that any measurement on a qubit in a superposition state will project it into a classical state[1]. As a result, in a recent work by Huang et al. [3], statistical assertions, meaning statistical analysis on multiple measurement results, are proposed to debug quantum programs. The key limitation of this approach is that each measurement stops the program execution and the assertions cannot be enabled when the actual computation results are to be measured. In this paper, we propose quantum circuits to overcome this limitation and to enable dynamic assertions for quantum programs.

Our proposed quantum circuits for dynamic assertions are inspired from quantum error correction. As qubits cannot be copied and cannot be measured directly, our approach for dynamic assertions is to indirectly verify the desired condition to be checked. In comparison, quantum error correction shares the same constraints and the various previously proposed quantum error correction codes [4] essentially introduce ancilla qubits and encode the information of the qubits to be protected in the ancilla qubits, which are checked and used to correct the qubits if they are corrupted. Similarly, we also introduce ancilla qubits for assertions but the difference is that we only need to check for assertions and our proposed quantum circuits for assertions are much simpler than those for error correction, which incurs very high overhead in the amount of ancilla qubits and the associated quantum circuits.

According to the previous work by Huang et al. [3], three types of possible assertions are essential for debugging quantum programs: classical assertions, superposition assertions, and entanglement assertions. Classical assertions check quantum variables with classical values to see whether they match the desired ones; superposition assertions check whether a quantum variable is in a desired superposition state; and entanglement assertions checks whether the entangled quantum variables exhibit the desired correlation. In this paper, we propose circuits for these three types of assertions respectively. Besides proving the correctness of our proposed designs, we also verify them on a quantum simulator and employ them on an actual quantum computer, IBM Q.

## 2 BACKGROUND AND RELATED WORK

Qubits are the foundation of quantum computing. Executing a quantum program means a sequence of operations upon the qubits. A qubit can be in a classical state, i.e., the $|0\rangle$ state or $|1\rangle$ state, which can be viewed as the classical 0 or 1 states. Besides classical states, a qubit can be in a superposition state, which is a linear combination of classical states, i.e., $|\psi\rangle = a|0\rangle+b|1\rangle$, where both $a$ and $b$ are complex number and $|a|^2+|b|^2=1$. When a qubit in the superposition state is measured, the superposition state is projected into a classical state with the probability of $|a|^2$ being state $|0\rangle$ and $|b|^2$ being state $|1\rangle$. Superposition states are the reason for quantum parallelism, as $n$ qubits can be in a mixture of $2^n$ states while in classical computing an $n$-bit variable takes *one* of the $2^n$ states at a time.

The state of multiple qubits can be described as the tensor product between the individual qubit state vectors. For example, the state of the two qubit, $|\psi\rangle = a|0\rangle+b|1\rangle$ and $|\delta\rangle = c|0\rangle+d|1\rangle$, can be described as $|\psi\rangle\otimes|\delta\rangle = ac|00\rangle + ad|01\rangle + bc|10\rangle+bd|11\rangle$, where $|00\rangle$ is $|0\rangle\otimes|0\rangle$ and $|01\rangle$ is $|0\rangle\otimes|1\rangle$, etc. Two or more qubits can be entangled, meaning that their measurements results should be always correlated. One implication is that among the entangled qubits, if one of them is measured (i.e., projected to a classical state), the rest will also collapse into a compatible classical state, losing their superposition states.

---

[1]. In this paper, "classical states" refer to a state in the computational basis, $|0\rangle$ and $|1\rangle$, and measurements are assumed to be performed with respect to the computational basis.

A quantum program is essentially a sequence of quantum gates performed upon a number of qubits. There are single-qubit gates such as Hadamard (H) gate, phase (S) gate, Pauli-X (X) gate, Pauli-Y (Y) gate, Pauli-Z (Z) gate, etc., and multi-qubit gates such as controlled-NOT (CNOT) gate. It has been proven that single-qubit gates and CNOT gates are universal for quantum computation. As we mainly use H gates and CNOT gates in this paper, we present their logic relationship in Figure 1.

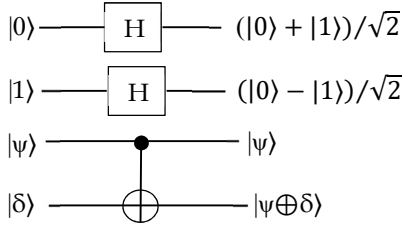

*Figure 1. Logic functions of the Hadamard gate and CNOT gate.*

Both superposition and entanglement are used extensively in quantum programs, and they are the fundamental reason for the computational advantage of quantum computing over classical computing. However, they do not have correspondence in classical computing, which makes them hard to reason about. The development of quantum programs remains a difficult task and debugging them is also very challenging. In the prior work, Huang et al. [3] analyzed a set of quantum programs and identified that the three following types of assertions are needed in quantum programs: assertions for classical values, assertions for superposition states, and assertions for entangled states. They proposed the statistical approach to realize these assertions by measuring the qubits many times. The limitation is that each measurement collapses the superposition state and projects the entangled qubits. As a result, such measurements stop the execution of the quantum program. When the execution is performed and the results are measured, such intermediate assertions could not be enforced. In the next section, we propose our quantum circuits to enable dynamic assertions, which can be checked when the quantum program is executed and the computation results are collected.

## 3 QUANTUM CIRCUITS FOR DYNAMIC ASSERTIONS

Inspired by quantum error correction, our key idea to enable dynamic assertion is to introduce additional quantum bits, aka ancilla qubits, to get information about the qubits under test, and to measure the ancilla qubits rather than directly measuring the qubits under test. This way, we do not need to disrupt the program execution when the assertion is checked. However, care needs to be taken to ensure that measuring the ancilla qubits will not affect the original quantum circuit. Next, we describe our proposed circuits for each type of assertion. For all the circuits, a measurement of the ancilla qubit being |1⟩ means an assertion error.

### 3.1 Dynamic Assertion for Classical Values

To ensure that the qubits are initialized to the correct values or some intermediate classical results should satisfy some conditions such as (|ψ⟩ != |0⟩), we can resort to assertions for classical values. We propose to introduce one ancilla qubit and a CNOT gate to achieve classical-value assertion for one qubit, as shown in Figure 2. In the figure, the qubit |ψ⟩ is to be checked for (|ψ⟩ ==|0⟩) . The ancilla qubit is initialized to |0⟩ and measured after the CNOT gate. If we initialize the ancilla qubit to be |1⟩, the same circuit asserts (|ψ⟩ ==|1⟩).

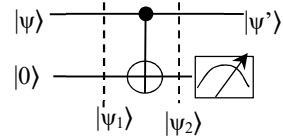

*Figure 2. Circuit for asserting classical values (|ψ⟩==|0⟩).*

**Proof.** In Figure 1, the state |ψ$_1$⟩ = |ψ⟩⊗|0⟩.
The state after the CNOT gate |ψ$_2$⟩ = |ψ⟩⊗|ψ⊕0⟩ = |ψ⟩⊗|ψ⟩.

If |ψ⟩ is in a classical state, either |0⟩ or |1⟩, then |ψ$_1$⟩ is either |00⟩ or |10⟩ and |ψ$_2$⟩ is |00⟩ or |11⟩, correspondingly. As a result, when the ancilla qubit is measured, if it is |0⟩, it means that |ψ⟩ must be |0⟩; if it is |1⟩, |ψ⟩ must be |1⟩, i.e., an assertion error.

If the qubit |ψ⟩ is in a superposition state due to a bug or a runtime error, |ψ⟩ = a|0⟩+b|1⟩. |ψ$_1$⟩ is a|00⟩+b|10⟩ and |ψ$_2$⟩ becomes a|00⟩+b|11⟩, which is actually an entangled state. Due to such entanglement, after the measurement of the ancilla qubit, if the measurement result is |0⟩ (i.e., no assertion error), the qubit under test will be projected into the classical state |0⟩, i.e, |ψ'⟩ = |0⟩. If the measurement result is |1⟩ (i.e., an assertion error), it is projected into the classical state, |1⟩. It means that when we perform an assertion check (|ψ⟩ ==|0⟩), if there is no assertion error, the proposed circuit may have automatically corrected the qubit if it is in a superposition state. If it cannot correct the qubit into the expected classical state, the assertion error occurs. Since the probability of the measurement result being |0⟩ and |1⟩ is $|a|^2$ and $|b|^2$, respectively, the probability distribution of assertion errors over repeated runs can be used to estimate a and b, if needed.

### 3.2 Dynamic Assertion for Entanglement

To assert that two or more qubits are in the entangled state of a|00⟩+b|11⟩ or a|01⟩+b|10⟩, we propose to leverage parity computation. Figure 3 shows the proposed circuit for computing the parity of two qubits. If checking whether the two qubits are entangled in the state of a|00⟩+b|11⟩, the ancilla qubit is initialized to |0⟩. If asserting that the two qubits are in the state of a|01⟩+b|10⟩, the ancilla should be initialized to |1⟩.

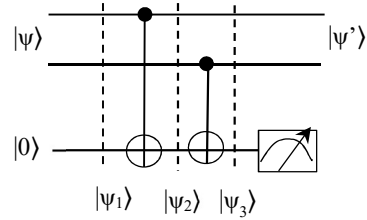

*Figure 3. Circuit for asserting entanglement.*

**Proof.** In Figure 3, if the input qubits are entangled in

the state of $a|00\rangle+b|11\rangle$, $|\psi\rangle = a|00\rangle+b|11\rangle$

Then, the state $|\psi_1\rangle = (a|00\rangle+b|11\rangle)\otimes|0\rangle = a|000\rangle+b|110\rangle$.

The state $|\psi_2\rangle = a|000\rangle+b|111\rangle$, i.e., the ancilla qubit is entangled as well after the CNOT gate.

The state $|\psi_3\rangle = a|000\rangle+b|110\rangle = (a|00\rangle+b|11\rangle)\otimes|0\rangle = |\psi\rangle\otimes|0\rangle$, which means that the ancilla qubit is un-entangled from the two qubits under test and should be $|0\rangle$. In addition, the qubits state $|\psi\rangle$ is unaffected for subsequent computations.

If the input qubits are not entangled, i.e., $|\psi\rangle = a|00\rangle+b|11\rangle+c|10\rangle+d|01\rangle$.

Then, $|\psi_1\rangle = a|000\rangle+b|110\rangle+c|100\rangle+d|010\rangle$.

$|\psi_2\rangle = a|000\rangle+b|111\rangle+c|101\rangle+d|010\rangle$.

$|\psi_3\rangle = a|000\rangle+b|110\rangle+c|101\rangle+d|011\rangle$, which means that when measuring the ancilla qubit, the result can be either $|0\rangle$ or $|1\rangle$. If $|0\rangle$, $|\psi_3\rangle$ is projected to $a'|000\rangle+b'|110\rangle = (a'|00\rangle+b'|11\rangle)\otimes|0\rangle$, i.e., the input qubits are forced into the entangled state. If $|1\rangle$, $|\psi_3\rangle$ is projected to $c'|101\rangle+d'|011\rangle = (c'|10\rangle+d'|01\rangle)\otimes|1\rangle$, i.e., another entangled state, while the assertion error is reported. The probability of measurement results being $|0\rangle$ or $|1\rangle$ can be used to compute the coefficients $a$, $b$, $c$, $d$, if needed.

Note that in Figure 3, the two CNOT gates act as inverse operation to each other when the qubits under test are entangled. Therefore, to assert more than two qubits (e.g., three) are entangled, we always need an even number of CNOT gates rather than the exact number of qubits. Figure 4 illustrates the case for asserting three entangled qubits. Otherwise, the ancilla qubit would remain entangled with the qubits under test, which would alter the functionality of subsequent computations.

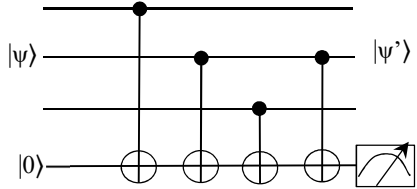

*Figure 4. Circuit for asserting three qubits are entangled.*

### 3.3 Dynamic Assertion for Superposition

In quantum computing, it is a common practice to use Hadamard gates to set the input qubits in the equal/uniform superposition state, $|+\rangle = 1/\sqrt{2}|0\rangle + 1/\sqrt{2}|1\rangle$, in order to take advantage of quantum parallelism. To assert such operations are correctly performed, we propose the circuit as shown in Figure 5.

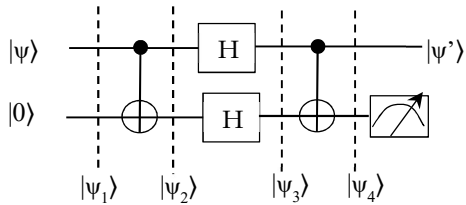

*Figure 5. Circuit for asserting equal superposition.*

**Proof.** In Figure 5, $|\psi\rangle = a|0\rangle+b|1\rangle$. If it is in the equal superposition state, i.e., $|\psi\rangle = |+\rangle$ or $a = b = 1/\sqrt{2}$.

The state $|\psi_1\rangle = (a|0\rangle+b|1\rangle)\otimes|0\rangle = a|00\rangle+b|10\rangle$.

The state $|\psi_2\rangle = a|00\rangle+b|11\rangle$.

The state $|\psi_3\rangle = a(|0\rangle + |1\rangle)/\sqrt{2}\otimes(|0\rangle + |1\rangle)/\sqrt{2} + b(|0\rangle - |1\rangle)/\sqrt{2}\otimes(|0\rangle - |1\rangle)/\sqrt{2}$
$= ½[(a|00\rangle + a|01\rangle + a|10\rangle + a|11\rangle) + (b|00\rangle - b|01\rangle - b|10\rangle + b|11\rangle)]$.

The state $|\psi_4\rangle = ½[(a|00\rangle + a|01\rangle + a|11\rangle + a|10\rangle) + (b|00\rangle - b|01\rangle - b|11\rangle + b|10\rangle)]$
$= ½[(a+b)|00\rangle+(a-b)|01\rangle+(a+b)|10\rangle+(a-b)|11\rangle]$.

If $|\psi\rangle = |+\rangle$ or $a = b = 1/\sqrt{2}$, then $|\psi_4\rangle = ½[(a+b)|00\rangle + (a+b)|10\rangle] = 1/\sqrt{2}[|00\rangle + |10\rangle] = |+\rangle\otimes|0\rangle$. This means that the ancilla qubit should always be $|0\rangle$ and it is un-entangled from the qubit under test. Therefore, the subsequent computation is not affected by the measurement of the ancilla qubit.

If $|\psi\rangle = |-\rangle$ or $a = 1/\sqrt{2}$ and $b = -1/\sqrt{2}$, then $|\psi_4\rangle = ½[(a-b)|01\rangle + (a-b)|11\rangle] = 1/\sqrt{2}[|01\rangle + |11\rangle] = |+\rangle\otimes|1\rangle$. This means that the ancilla qubit should always be $|1\rangle$ and it is un-entangled from the qubit under test.

If $|\psi\rangle \ne |+\rangle$ or $|-\rangle$, then the ancilla qubit and the qubit under test remain entangled. When the ancilla qubit is measured, the state $|\psi_4\rangle$ will be projected. The probability of the measurement result being $|0\rangle$ is the probability of the state $|\psi_4\rangle$ being in the state of $|00\rangle$ or $|10\rangle$. Therefore, the probability can be computed as $[|a+b|^2 + |a+b|^2] / [|a+b|^2 + |a-b|^2+ |a+b|^2+|a-b|^2]$. If both $a$ and $b$ are real, then the probability becomes $(2a^2+4ab+2b^2)/4 = (2 + 4ab)/4$. Similarly, we can derive the probability of the measurement result on the ancilla qubit being $|1\rangle$ as $[|a-b|^2 + |a-b|^2] / [|a+b|^2 + |a-b|^2 + |a+b|^2+|a-b|^2]$, which becomes $(2a^2+4ab+2b^2)/4 = (2 - 4ab)/4$ if both $a$ and $b$ are real. In the case of $|\psi\rangle$ being in a classical state, i.e., $a = 0$ and $b = 1$ or $a = 1$ and $b = 0$, the measurement result on the ancilla qubit has the equal probability of 50% being $|0\rangle$ or $|1\rangle$.

Now, let us check the impact of the ancilla qubit measurement on the qubit under test in this case. If the measurement result on the ancilla qubit is $|0\rangle$, i.e., no assertion error, $|\psi_4\rangle$ is projected to: $½[(a'|00\rangle + a'|10\rangle) + (b'|00\rangle + b'|10\rangle)] = ½[(a'+b')|00\rangle + (a'+b')|10\rangle]$
$= ½[(a'+b')|0\rangle + (a'+b')|1\rangle]\otimes|0\rangle$
$= |\psi'\rangle\otimes|0\rangle$.

On the other hand, if the measurement result of the ancilla qubit being $|1\rangle$, $|\psi_4\rangle$ is projected to: $½[(a'|01\rangle + a'|11\rangle) - (b'|01\rangle + b'|11\rangle)] = ½[(a'-b')|0\rangle + (a'-b')|1\rangle]\otimes|1\rangle$
$= |\psi'\rangle\otimes|1\rangle$

In both cases, as the coefficients of $|0\rangle$ and $|1\rangle$ of $|\psi'\rangle$ are identical and they must satisfy the unitary condition, the magnitude of the coefficient must be $1/\sqrt{2}$. In other words, in the case of $|\psi\rangle \ne |+\rangle$, no matter whether the measurement result of the ancilla qubit being $|0\rangle$ or $|1\rangle$, the qubit after the assertion circuit is forced into the superposition state, $|\psi'\rangle = k|0\rangle + k|1\rangle$, where $|k| = 1/\sqrt{2}$.

As discussed earlier, the probabilities of the measurement result of the ancilla qubit being $|0\rangle$ or $|1\rangle$ can be used to compute the magnitude of the original coefficients $a$ and $b$. The equal probability of the measurement result being $|0\rangle$ or $|1\rangle$ indicates that the qubit under test is likely in a classical state.

## 4 EXPERIMENTS

We conduct experiments on the quantum circuit simulator, QUIRK [1], to verify our mathematical derivation and an IBM Q (ibmqx4) quantum computer to check the effectiveness of the assertions in filtering errorneous results. The use of assertions for debugging harnesses is discussed in detail by Huang et al. [3].

### 4.1 Classical Assertions

We first construct the circuit in QUIRK and set the input as classical values to verify both the measurement results on the ancilla qubit (i.e., the assertion result) and the state of the qubit under test. Then, we set the input to a superpositoned state and use assertion measurement to project the qubit under test, as shown in Figure 6. To simulate the projection effect, we add a post-select operator, which ignores the result when there is an assertion error. As shown in the figure, the input qubit, which is in the superposition state, is forced to be |0⟩ after the assertion check.

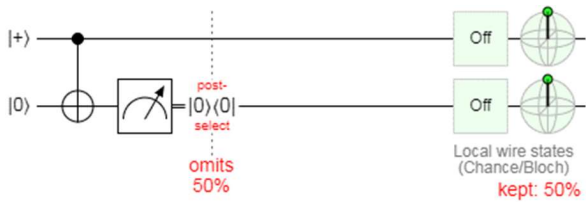

*Figure 6. Verifying the classical assertion circuit using QUIRK. Although the input qubit is |+⟩, it is projected to |0⟩ as a result of the measurement of the ancila qubit.*

We implemented the circuit on the 5-qubit IBM Q quantum computer. Due to the constraints on connectivity of the IBM Q computer, we used qubit $q_2$ as the ancilla qubit to assert the qubit ($q_1$ == |0⟩). The experimental results are shown in Table 1. Based on the statistics, we can see that if we do not use the assertion check, the error rate is 3.5% (=2.4%+1.1%) where $q_1$ is |1⟩ although we expected it is to be |1⟩. If we discard all the results with an assertion error, the error rate is reduced to 2.4%/(93.8%+2.4%) = 2.5%, a reduction of 28.5% in the error rate.

*Table 1. The results of classical assertion logic on IBM Q.*

| $q_1q_2$ | % | Meaning |
|---|---|---|
| 00 | 93.8% | No assertion error, q1 is 0 |
| 01 | 2.7% | Assertion error, q1 is 0 (potential false positive) |
| 10 | 2.4% | No assertion error, q1 is 1 (false negative) |
| 11 | 1.1% | Assertion error, q1 is 1. |

### 4.2 Entanglement Assertions

We use the same approach to verify our circuits for entanglement assertions using QUIRK. Although the details are omitted due to the limited space, the simulation results confirm the correctness of our derivation in Section 3.2.

On the 5-qubit IBM Q quantum computer, we used an H gate and a CNOT gate to entangle the qubit q1 and q2 into the uniform superposition state (|00⟩+|11⟩). Then, we use qubit q0 as the ancilla qubit to assert the entanglement between q1 and q2. The experimental results are shown in Table 2. From the statistics, we can see that the error rate of the expected entanglement state is 18.4% (=6.3%+4.4%+5.6%+2.1%). If we filter out the ones with assertion errors, the error rate is reduced to (6.3%+4.4%)/(39.1%+6.3%+4.4%+34.6%) = 12.6%, an improvement of 31.5%.

*Table 2. The results of entanglement assertion logic on IBM Q.*

| $q_0q_1q_2$ | % | Meaning |
|---|---|---|
| 000 | 39.1% | No assertion error, q1 q2 entangled |
| 001 | 6.3% | No assertion error, q1 q2 not entangled (false negative) |
| 010 | 4.4% | No assertion error, q1q2 not entangled (false negative) |
| 011 | 34.6% | No assertion error, q1 q2 entangeld |
| 100 | 4.0% | Assertion error (potential false positive) |
| 101 | 5.6% | Assertion error, q1 q2 not entangled |
| 110 | 2.1% | Assertion error, q1 q2 not entangled |
| 111 | 3.9% | Assertion error (potential false positive) |

### 4.3 Superposition Assertions

Using QUIRK, we explore inputs with different states to verify our derivations in Section 3.3. In Figure 7, we show when the input is in a classical state, the qubit is forced into the superposition state after the measurement on the ancilla qubit. The measurement also indicates a 50% assertion error rate, confirming our derivation in Section 3.3.

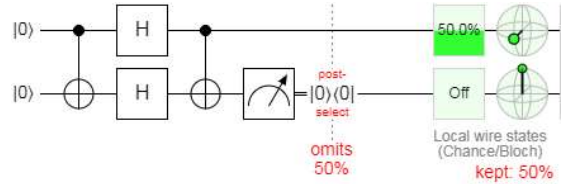

*Figure 7. Verifying the superposition assertion circuit using QUIRK. The input is set to a classical state and it becomes the superposition state after the assertion logic.*

We also implement the superposition assertion circuit on the IBM Q quantum computer. Since the qubit under test is supposed in a uniform superposition state, the measurement result can be 0 or 1. Therefore, it is not obvious to tell from the measurement result, either being 0 or 1, whether there should be an assertion error or not. Anyway, our proposed assertion circuit reports an assertion error in 15.6% of the measurements, indicating that it captures some erroneous effect of the expected superposition state.

## 7 CONCLUSIONS

In this paper, we propose quantum circuits to enable dynamic assertions for classical values, entanglement, and superposition. We prove the functions of the proposed circuits and verify them with a quantum simulator. We show that besides generating assertion errors, the assertion logic may also force the qubits under test to be into the desired state. Our proposed assertion logic can also be used in the noisy intermediate scale quantum (NISQ) systems to filter out errionous results, as exemplified on the 5-qubit IBM Q quantum computer.

### REFERENCES

[1] QUIRK, a quantum circuit simulator. https://algassert.com/quirk

[2] Yipeng Huang and Margaret Martonosi, "QDB: From Quantum Algorithms Towards Correct Quantum Programs", PLATEAU Workshop at ACM conference on Systems, Programming, Languages and Applications: Software for Humanity (SPLASH). 2018.

[3] Yipeng Huang and Margaret Martonosi, "Statistical Assertions for Validating Patterns and Finding Bugs in Quantum Programs", ISCA'19.

[4] Michael A. Nielsen and Isaac L. Chuang, "Quantum Computation and Quantum Inofmration", Cambridge Press, 2000.